\documentclass[symmetry,article,submit,moreauthors,pdflatex,10pt,a4paper]{mdpi} 
\usepackage{graphicx}
\usepackage{youngtab,bbm} 
\preto{\abstractkeywords}{\nolinenumbers}
\firstpage{1} 
\articlenumber{x}
\doinum{10.3390/------}
\pubvolume{xx}
\pubyear{2018}
\copyrightyear{2018}
\externaleditor{Academic Editor: Charles D.\ Lane}
\history{Received: August 31, 2018; Accepted: October 9, 2018; Published: date}

\pdfoutput=1

\newcommand*\al{\alpha}
\newcommand*\be{\beta}
\newcommand*\ga{\gamma}
\newcommand*\de{\delta} 
\newcommand*\ep{\epsilon}

\newcommand*\et{\eta}

\newcommand*\ka{\kappa}
\newcommand*\la{\lambda} 

\newcommand*\rh{\rho}

\newcommand*\si{\sigma}

\newcommand*\om{\omega}
\newcommand*\Ga{\Gamma}

\newcommand*\La{\Lambda}

\newcommand*\mn{{\mu\nu}} 

\newcommand*\pt[1]{\phantom{#1}}
\newcommand*\prt{\partial}

\newcommand*\fr[2]{{\textstyle{{#1} \over {#2}}}}

\newcommand*\lsim{\mathrel{\rlap{\lower4pt\hbox{\hskip1pt$\sim$}}
    \raise1pt\hbox{$<$}}}
\newcommand*\gsim{\mathrel{\rlap{\lower4pt\hbox{\hskip1pt$\sim$}}
    \raise1pt\hbox{$>$}}}
\newcommand*\sqr[2]{{\vcenter{\vbox{\hrule height.#2pt
         \hbox{\vrule width.#2pt height#1pt \kern#1pt
         \vrule width.#2pt}
         \hrule height.#2pt}}}}

\newcommand*\hb{\hbar}

\newcommand*\x{{\hat x}}

\newcommand*\etal{{\it et al.}}

\newcommand*\labeleeq[1]{ \label{#1} \end{equation} } 
\newcommand*\labeleea[1]{ \label{#1} \end{eqnarray} }

\newcommand{\beq}{\begin{equation}}
\newcommand{\eeq}{\end{equation}}
\newcommand{\bea}{\begin{eqnarray}}
\newcommand{\eea}{\end{eqnarray}}
\newcommand{\rf}[1]{Eq.\ (\ref{#1})}

\newcommand{\eq}[1]{Eq.\ (\ref{#1})}
\newcommand{\Eq}[1]{Eq.\ (\ref{#1})}

\newcommand{\altnabla}{\widetilde{\nabla}} 

\newcommand{\alttextcolor}[2]{\textcolor{black}{#2}}

\Title{Relating Noncommutative SO(2,3)$_\star$ Gravity 
to the Lorentz-Violating Standard-Model Extension}

\Author{Quentin G.\ Bailey $^{1}$ 
 and Charles D.\ Lane $^{2}$*\orcidA{}}

\AuthorNames{Quentin G.\ Bailey and Charles D.\ Lane}

\address{%
$^{1}$ \quad Embry-Riddle Aeronautical University, Prescott, AZ; baileyq@erau.edu\\ 
$^{2}$ \quad Berry College and Indiana University Center for Spacetime Symmetries; clane@berry.edu} 

\corres{Correspondence: clane@berry.edu}

\abstract{We consider a model of noncommutative gravity 
that is based on a spacetime with broken local SO(2,3)$_\star$ symmetry. 
We show that the torsion-free version of this model 
\alttextcolor{red}{
is} 
contained within 
the framework of the Lorentz-violating Standard-Model Extension. 
\alttextcolor{red}{ 
We analyze in detail the relation between the torsion-free, quadratic limits 
of the broken SO(2,3)$_\star$ model and the Standard-Model Extension. 
} 
As part of the analysis, 
we construct the relevant geometric quantities to quadratic order 
in the metric perturbation around a flat background.}


\begin{document}

\section{Introduction}

While noncommutative geometry has been studied for more than 70 years 
\cite{Snyder:1946qz}, 
it has been especially popular as
\alttextcolor{red}{a} 
possible framework for physics beyond the Standard Model 
in recent decades 
\cite{Connes:1994yd,Seiberg:1999vs}. 
In particular, 
several extensions to general relativity 
that incorporate noncommutative geometry have been proposed 
\cite{Aschieri:2005yw, Aschieri:2005zs, Ohl:2009pv, Yang:2006dk, 
 Steinacker:2010rh, Chamseddine:2000si, Cardella:2002pb}. 
In this paper, we consider one particular model 
that is based on a flat spacetime with broken SO(2,3)$_\star$ symmetry 
\cite{Dimitrijevic:2014,Ciric:2016isg}. 

Any physical model that includes noncommutative effects 
and that reduces to conventional physics in the proper limit 
is expected to break Lorentz symmetry \cite{Carroll:2001ws}. 
A general framework for 
\alttextcolor{red}{the} study of Lorentz violation 
has been developed over the last 
\alttextcolor{red}{30} years 
\cite{Kostelecky:1988zi,Kostelecky:1994rn,Colladay:1996iz,Colladay:1998fq,Kostelecky:2003fs}. 
Indeed, 
numerous experimental and observational limits exist already on many different {\it a priori}
independent types of Lorentz violation \cite{Kostelecky:2008ts}.
Additionally, 
this effective-field-theory framework should contain any realistic noncommutative model. 
This has already been shown for non-gravitational models \cite{Carroll:2001ws}. 
In this work, 
we argue that the noncommutative SO(2,3)$_\star$ gravity model also fits into 
the gravitational sector of the Standard-Model Extension. 
\alttextcolor{red}{ 
This serves as an example of the general notion 
that the SME contains all specific \alttextcolor{red}{action-based} Lorentz-violating models. 
} 

\section{Noncommutative SO(2,3)$_\star$ Gravity} 

Consider a model consisting of a flat 4-dimensional spacetime 
with an SO(2,3) gauge field \cite{Ciric:2016isg}. 
Suppose that this symmetry is spontaneously broken along a timelike direction, 
with the field in that direction achieving a vacuum expectation value $\ell$. 
The corresponding action takes the form of a model of gravity, 
with pieces corresponding to Einstein-Hilbert terms, 
cosmological-constant terms, and Gauss-Bonnet terms; 
\alttextcolor{red}{this action is symmetric under an SO(1,3) subgroup 
of the broken SO(2,3) symmetry.} 
If conventional field products are then translated into Moyal-Weyl $\star$-products 
and a Seiberg-Witten map is used to re-express quantities 
in terms of commutative products, 
we get a broken-SO(2,3)$_\star$ gravitational theory. 

\alttextcolor{red}{This process has been carried out in Ref.\ \cite{Ciric:2016isg} and we present the relevant results here. }
This theory may be expressed as a model with noncommutative local SO(1,3)$_\star$ symmetry. 
\alttextcolor{red}{The result is expanded in terms of the noncommutative background $\theta^{\al\be}$, 
with leading terms at second order in this quantity.
To display the action, 
we note that the geometric quantities that will appear use conventional notation: 
${e_\mu}^a$ is the vierbein (with determinant $e$), 
${\om_\ga}^{ab}$ is the associated spin connection, 
${\Ga^\rh}_{\ga\al}$ are the Christoffel symbols 
associated with spacetime metric $g_{\al\be}$, 
$R_{\al\be\ga\de}$ is the Riemann tensor, 
$R_{\al\be}$ is the Ricci tensor, 
and $R$ is the curvature scalar.} 

\alttextcolor{red}{Once spacetime torsion $T_{\la\mu\nu}$ is set to zero, 
the action for the model (\cite{Ciric:2016isg}, Eq.\ (4.2)) 
may be expressed in the form 
}
\bea 
S_{NCR} &=& 
 - \frac{1}{2 \ka} \int d^4x\ 
 e\left[ R - \frac{6}{\ell^2}(1+c_2+2c_3) \right] 
 \nonumber \\ 
&& + \frac{1}{16 \ka \ell^4} \int d^4x \sum_{u=1}^{6} e \theta^{\al\be} \theta^{\ga\de}  C_{(u)} L^{(u)}_{\al\be\ga\de} 
 \quad,
 \label{SNCR} 
\eea 
where $\ka=8\pi G_N$ and $\ell$ is a length parameter.  
The antisymmetric coefficients $\theta^{\al\be}$ are to be thought of 
as a fixed background field describing the degree of noncommutativity of spacetime.
Note that natural units are adopted ($\hb=c=1$) which implies
that $\ell$ has units of length or inverse mass and $\theta$ has units of length squared. 

The top row of \alttextcolor{red}{\eq{SNCR}} 
is the action for conventional general relativity 
with a cosmological constant 
$\La = -3\left(\frac{1+c_2+2c_3}{\ell^2}\right)$. 
(Note that this is the correct value of the cosmological constant 
only in the commutative limit $\theta=0$. 
For $\theta\ne 0$, 
other terms in the action will also effectively contribute to it.) 
The parameters $c_2$ and $c_3$ describe the relative weights 
of various contributions to the unbroken SO(2,3)$_\star$ action. 
Thus, the action $S_{CNR}$ may be thought of as a family of actions 
parameterized by $c_2$, $c_3$, and $\ell$. 
The tensors $L^{(u)}_{\al\be\ga\de}$ are geometric quantities; 
the weights $C_{(u)}$ measure the relative contributions of these quantities 
to the action. 
The tensors and their weights are listed in Table 1. 

\begin{table}[H] 
\label{actionterms} 
\caption{Geometric quantities and their weights 
that appear in the action $S_{CNR}$.}
\centering
\[ \begin{array}{ccc} 
u  & \mbox{Weight }C_{(u)} & \mbox{Geometric Quantity }L^{(u)}_{\al\be\ga\de} \\ 
1  & 3 c_2+16 c_3        & R_{\al\be\ga\de} \\ 
2  & -6-22c_2-36c_3        & g_{\be\de} R_{\al\ga} \\ 
3  & \fr{1}{\ell^2}(6+28c_2+56c_3) & g_{\al\ga}g_{\be\de} \\ 
4  & -4-16c_2-32c_3        & e^\mu_a e_{\be b} (\altnabla_\ga e^a_\al) (\altnabla_\de e^b_\mu) \\ 
5  & 4+12c_2+32c_3         & e_{\de a} e^\mu_b (\altnabla_\al e^a_\ga) (\altnabla_\be e^b_\mu) \\ 
6  & 2+4c_2+8c_3           & g_{\be\de} e^\mu_a e^\nu_b [ (\altnabla_\al e^a_\nu) (\altnabla_\ga e^b_\mu) - (\altnabla_\ga e^a_\mu) (\altnabla_\al e^b_\nu) ] \\
\end{array} \] 
\end{table}

The adjusted covariant derivatives $\altnabla_\mu$ of the vierbein 
that appear in terms 
\alttextcolor{red}{4} through \alttextcolor{red}{6}  
include contributions 
from the SO(1,3) connection but not from the Christoffel symbols: 
\beq 
\altnabla_\ga {e_\al}^a = \prt_\al {e_\al}^a + {\om_\ga}^{ab}e_{ab} 
 = \nabla_\ga {e_\al}^a + {\Ga^\rh}_{\ga\al}{e_\rh}^a 
 \quad . 
\eeq 
If the vierbein satisfies the usual compatibility condition 
$\nabla_\ga {e_\al}^a=0$, 
then the adjusted covariant derivative may be expressed as 
\beq 
\altnabla_\ga {e_\al}^a = {\Ga^\rh}_{\ga\al}{e_\rh}^a 
 \quad. 
\eeq 
This implies the explicit appearance of the Christoffel symbols in the lagrangian, 
the consequences of which are discussed in the next section.

The model acts like a relativistic theory of gravity in several ways, 
but there are some issues with interpreting it as such. 
For example, 
it is derived with the assumption that $\prt_\al \theta^{\mu\nu}=0$. 
This assumption is reasonable in the original flat-spacetime context 
of the model. 
However, 
if the model is to be interpreted \alttextcolor{red}{in} curved spacetime, 
this assumption is clearly coordinate dependent. 
We may attempt to fix this issue by instead assuming that $\nabla_\al \theta^{\mu\nu}=0$, 
but even this condition cannot apply in many situations. 
Nonzero tensor fields with vanishing covariant derivative cannot exist on many manifolds, 
including, say, spacetime with a Schwarzschild metric 
\cite{Kostelecky:2003fs,Lane:2016osk}. 
Therefore, 
if we wish to seriously consider action \ref{SNCR} 
to represent a theory of gravity, 
then we must consider it 
to be an approximation to a more realistic model 
with $\nabla_\al \theta^{\mu\nu}\ne 0$. 
In what follows, 
we will assume that terms involving derivatives of $\theta^{\mu\nu}$ 
that may appear in a more-realistic model 
are negligible in comparison to all other terms. 

\section{Gravitational Sector of the Lorentz-Violating Standard-Model Extension} 

\alttextcolor{red}{ 
The full action \cite{Kostelecky:2003fs} describing the gravitational sector of the SME 
can be expressed as a sum of terms, 
each of which contracts a coefficient with spacetime indices 
with geometric quantities 
such as the Riemann tensor $R_{\al\be\ga\de}$, 
the torsion $T_{\la\mu\nu}$, and their covariant derivatives: 
\beq 
S_{\rm gravity} = \frac{1}{2\ka} \int d^4x\, e \left[ 
(k_T)^{\la\mu\nu} T_{\la\mu\nu} 
+(k_R)^{\ka\la\mu\nu} R_{\ka\la\mu\nu} 
+(k_{DT})^{\ka\la\mu\nu} D_\ka T_{\la\mu\nu} 
+ \cdots 
\right] 
\quad . 
\eeq 
The tensors $k_T$, $k_R$, etc.\ are coefficients for Lorentz and diffeomorphism violation
and the ellipses represent terms with higher powers of curvature and torsion and derivative terms \cite{Bailey:2015, Bailey:2016}.
Note that a violation of local Lorentz symmetry generically implies a violation of diffeomorphism symmetry, 
as explained in the literature \cite{Kostelecky:2003fs,Bluhm:2005}.
As with $\theta^{\mu\nu}$, 
it is not possible for the coefficients to be covariant derivative constants on most spacetime manifolds, 
and so they must be functions of spacetime position, 
though we may assume that their partial derivatives are negligible 
in experimentally relevant frames. 
}

\alttextcolor{red}{ 
In this work, 
we consider two limits of this full action: 
the minimal set of terms necessary for Lorentz violation 
and the weakly-curved-spacetime limit (or quadratic limit) of the full action. 
} 

\subsection{Covariant Match} 

In the gravity sector of the fully observer-covariant SME, 
the minimal set of terms that arises are given by the action \cite{Kostelecky:2003fs},
\beq 
S_{LV, {\rm cov}} = 
 \frac {1}{2\ka} \int d^4x \, e \left[ R + (k_R)_{\al\be\ga\de} R^{\al\be\ga\de} \right],
\label{Scov} 
\eeq 
where $(k_R)_{\al\be\ga\de}$ are the 20 (background) coefficients for local Lorentz and diffeomorphism violation.
It is clear that there is overlap with the noncommutative model \rf{SNCR}.
However, 
there are no terms in the SME containing explicit dependence on the non-tensorial connection coefficients
$\Ga^\al_{\pt{\al}\be\ga}$.

It is important at this stage to distinguish two types of symmetry transformations.
The first is called an {\it observer} diffeomorphism, or general coordinate transformation, 
which is a diffeomorphism that affects both the background, $(k_R)_{\al\be\ga\de}$ 
and the dynamical fields ${e_\mu}^a$. 
The second is called a {\it particle} diffeomorphism, 
which is a diffeomorphism that leaves the background $(k_R)_{\al\be\ga\de}$ unchanged 
while the dynamical fields ${e_\mu}^a$ transform in the usual way.
It is this second type of symmetry breaking, 
{\it particle} diffeomorphism symmetry breaking, 
that is described by the SME approach
and is broken by the second term in \rf{Scov}.
Because the action terms in the SME are scalars under general coordinate transformations, 
they trivially satisfy observer symmetry.
These points are discussed in more detail 
in the literature \cite{Kostelecky:2003fs, Bluhm2008, Bluhm2015}.

The explicit appearance of ${\Ga^\rh}_{\ga\al}$ in terms 4--6 of \rf{SNCR} implies that each of these terms 
is not symmetric under observer diffeomorphisms.
Whether the model can be massaged into an observer covariant form, 
\alttextcolor{red}{for example by a special choice of the parameters $c_2$ and $c_3$,}
remains to be shown.
Note that the model does appear covariant under \alttextcolor{red}{{\it observer}} local Lorentz transformations
while breaking \alttextcolor{red}{{\it particle}} local Lorentz symmetry.  
Despite the difficulty, 
we can proceed at the quadratic-action level, 
where a model that breaks observer diffeomorphism invariance cannot be distinguished 
from a model that breaks particle diffeomorphism invariance.

\subsection{Linearized Lorentz-Violating Standard-Model Extension} 
If we restrict the full SME to a version with equations of motion 
that are linear in $h_{\mu\nu}$ \cite{Kostelecky:2015,Kostelecky:2016,Kostelecky:2017zob}, 
then the action takes the form, 
after a rescaling by $1/2\ka$, 
\beq 
S= 
 \int d^4x \left[ {\cal L}_0 
 + \fr{1}{8\ka}h_{\mu\nu} \sum_d \widehat{\cal K}^{(d)\mu\nu\rh\si}h_{\rh\si} \right] 
 \quad. 
\label{SLV} 
\eeq 
In this expression, 
${\cal L}_0=e(R-2\La)/2\ka$ is the usual quadratic Einstein-Hilbert Lagrange density 
and $h_{\mu\nu}:=g_{\mu\nu}-\et_{\mu\nu}$ is the metric perturbation, 
assumed to be small. 
The $\widehat{\cal K}^{(d)\mu\nu\rh\si}$ are general derivative operators
formed from background coefficients and derivatives.
The summation is over the mass dimension $d$ of the operators.
In general, 
apart \alttextcolor{red}{from} surface terms, 
this sum includes 14 classes of irreducible representations 
involving tensors and derivative operators, 
all detailed in \cite{Kostelecky:2017zob}.

A primary goal of this paper is to argue that the non-commutative (broken-)SO(2,3)$_\star$ 
action $S_{NCR}$ in the linearized limit is a special case 
of this general linearized Lorentz-violating action. 
We explicitly calculate the map that shows this correspondence. 
We will show that the subset of the operator terms in the action \rf{SLV} that occur 
in the non-commutative model \rf{SNCR} can be written as
\bea 
S_{LV, NC} &=& 
\fr{1}{8\ka} \int d^4x \, h_{\mu\nu} \big\{  
\big[ s^{(4)\mu\rh\al\nu\si\be}+s^{(4,1)\mu\rh\nu\si\al\be} +s^{(4,2)\mu\rh\al\nu\si\be} + k^{(4,3)\mu\al\nu\be\rh\si} \big] \prt_\al \prt_\be
\nonumber\\
&& \pt{needspace+}
 +s^{(2,1)\mu\rh\nu\si} + k^{(2,1)\mu\nu\rh\si}
\big\}  h_{\rh\si}.
\label{SLVNC} 
\eea 
Each of these terms has distinct tensor symmetries \alttextcolor{red}{described} 
by a particular Young tableau \cite{hamermesh}.
The coefficients with the $(4,\#)$ label are coefficients for mass dimension $4$ operators, 
while those without derivatives labeled $(2,\#)$ are coefficients for mass dimension $2$ operators.
The latter represent an arbitrary mass matrix for the gravitational fluctuations $h_{\mu\nu}$.
Incidently, 
none of the terms in \rf{SNCR} contain odd mass dimension operators, 
and therefore the $CPT$ symmetry is maintained.

\section{Connecting NC SO(3,2)$_\star$ Gravity to the SME} 

The action for any linearized theory of gravity 
is quadratic in the perturbation $h_{\mu\nu}$. 
Therefore, we need to calculate each of the quantities that appears in $S_{NCR}$ 
to second order in $h_{\mu\nu}$. 
Calculations of these quantities to first order are widespread in the literature, 
but calculations to second order are not, 
so we summarize the key results in the Appendix. 
With these formul\ae, 
we may expand the noncommutative action $S_{NCR}$ in powers of $h_{\mu\nu}$. 
The results may then be \alttextcolor{red}{manipulated} into the form of the linearized action \Eq{SLV}. 

First, 
we show the match to the SME for the {\it massive} $u=3$ term.
Expanding this term from \rf{SNCR} in the quadratic action limit, 
we obtain
\bea 
S_{{\rm NC, Mass}} &=& 
\frac{C_{(3)} }{16 \ka \ell^6} \int d^4x e \theta^{\al\be} \theta^{\ga\de}  g_{\al\ga} g_{\be\de}
\nonumber\\
&=& \frac{1}{8 \ka} \int d^4x \Big\{
\fr{C_{(3)} }{2 \ell^6} \left( \theta^2+\left[ \fr{1}{2}\theta^2 \et^{\mu\nu} 
+ 2{\theta_\al}^\mu \theta^{\al\mu} \right] h_{\mu\nu} \right)
  \nonumber \\ 
 && \pt{need space+}+ \fr{C_{(3)} }{16 \ell^4} h_{\mu\nu} \left[ 
  \theta^2 \et^{\mu\nu}\et^{\rh\si} 
  -2\theta^2 \et^{\mu\rh}\et^{\nu\si} 
  +8 {\theta_\al}^\mu \theta^{\al\nu} \et^{\rh\si} 
  +8 \theta^{\mu\rh} \theta^{\nu\si} 
  \right] h_{\rh\si} \Big\}, 
\eea 
where $\theta^2 := \theta_{\mu\nu}\theta^{\mu\nu}$. 
Note that all indices on the right-hand sides of these expressions 
are raised and lowered with $\et$, 
as they are considered to act in the flat spacetime with field $h_{\mu\nu}$. 
The first term with just $\theta^2$ is a constant and irrelevant for dynamics, 
while the second term linear in $h_{\mu\nu}$ acts as a constant contribution to the stress-energy tensor (of the form 
of a cosmological constant). 
The last line can be matched to the last two terms in \rf{SLVNC} using Young tableau projections.
The coefficients appearing, 
$s^{(2,1)\mu\rh\nu\si}$ and $k^{(2,1)\mu\nu\rh\si}$, correspond to the 
Young tableaus $\scriptsize\young(\mu\nu,\rh\si)$ and $\scriptsize\young(\mu\nu\rh\si)$, 
respectively.
The explicit results we find are 
\bea
s^{(2,1)\mu\rh\nu\si} &=& \frac{C_{(3)} }{12 \ell^4} 
\big[ 2\et^{\mu\nu} \theta^{\rh\al} \theta^{\si}_{\pt{\si}\al}+2\et^{\rh\si} \theta^{\mu\al} \theta^{\nu}_{\pt{\si}\al}
-2\et^{\rh\nu} \theta^{\si\al} \theta^{\mu}_{\pt{\mu}\al} -  2\et^{\mu\si} \theta^{\rh\al} \theta^{\nu}_{\pt{\si}\al}
\nonumber\\
&&
\pt{space}+ 2\theta^{\rh\nu} \theta^{\si\mu}+ 4 \theta^{\rh\mu} \theta^{\si\nu}+ 2\theta^{\mu\nu} \theta^{\rh\si}
+ \left(\et^{\rh\si} \et^{\mu\nu} -\et^{\rh\nu} \et^{\si\mu} \right)\theta^2 
\big] 
\quad , 
\nonumber\\
k^{(2,1)\mu\nu\rh\si} &=& \frac{C_{(3)} }{48 \ell^4} 
\big[ 4\et^{\mu\nu} \theta^{\rh\al} \theta^{\si}_{\pt{\si}\al} + 4\et^{\rh\si} \theta^{\mu\al} \theta^{\nu}_{\pt{\si}\al}
+4\et^{\rh\nu} \theta^{\si\al} \theta^{\mu}_{\pt{\mu}\al} + 4\et^{\mu\si} \theta^{\rh\al} \theta^{\nu}_{\pt{\si}\al}
\nonumber\\
&&
+4\et^{\si\nu} \theta^{\rh\al} \theta^{\mu}_{\pt{\mu}\al} + 4\et^{\rh\mu} \theta^{\si\al} \theta^{\nu}_{\pt{\si}\al}
- \left(\et^{\rh\si} \et^{\mu\nu} +\et^{\rh\nu} \et^{\si\mu} +\et^{\rh\mu} \et^{\si\nu}\right)\theta^2 
\big] 
\quad . 
\label{mass}
\eea

We classify the remaining terms in \rf{SNCR} as 
{\it kinetic} terms that only involve mass dimension $4$ operators.
After expanding these terms in the quadratic-action limit and manipulating the result into the form of \rf{SLVNC}, 
we obtain
\bea 
S_{{\rm NC,Kin}} &=& \frac{1}{8 \ka} \int d^4x \, 
h_{\mu\nu} (K_{\rm NC})^{\mu\nu\rh\si\al\be} \prt_\al \prt_\be h_{\rh\si},  
\label{NCKin}
\eea 
where the quantity $(K_{\rm NC})^{\mu\nu\rh\si\al\be}$ is given by
\bea
(K_{\rm NC})^{\mu\nu\rh\si\al\be} &=& 
\fr {1}{16 \ell^4} (2 C_{(1)} - 2 C_{(2)} + C_{(4)} ) ( \et^{\al\be} \theta^{\rh\nu} \theta^{\si\mu} 
+ \et^{\al\be} \theta^{\rh\mu} \theta^{\si\nu} )
\nonumber\\
&& +\fr {1}{64\ell^4} (4 C_{(1)} -2 C_{(2)} + C_{(4)} ) 
\big( \{ ( \et^{\nu\al} \theta^{\be\rh} \theta^{\si\mu}-\et^{\si \al} \theta^{\be\mu} \theta^{\rh\nu}
+\et^{\nu\be} \theta^{\al\rh} \theta^{\si\mu}-\et^{\si \be} \theta^{\al\mu} \theta^{\rh\nu})
\nonumber\\
&&
\pt{bigggspace much much more ()} +(\rh  \rightleftharpoons \si ) \} + \{ \mu  \rightleftharpoons \nu \} 
\big)
\nonumber\\
&&
+\fr {1}{16 \ell^4} (2 C_{(1)}+ C_{(2)} - C_{(5)}) \big( \et^{\mu\nu} \theta^{\rh\al} \theta^{\be\si} 
+ \et^{\rh\si} \theta^{\mu\al} \theta^{\be\nu} + \et^{\mu\nu} \theta^{\rh\be} \theta^{\al\si} 
+ \et^{\rh\si} \theta^{\mu\be} \theta^{\al\nu} \big)
\nonumber\\
&&
+\fr {1}{16 \ell^4} (C_{(2)} - C_{(6)})\big( \{ 
(  \fr 12 \et^{\si\al} \et^{\be\nu} \theta^{\rh\ga} \theta^{\mu}_{\pt{\mu}\ga} 
+ \fr 12 \et^{\si\be} \et^{\al\nu} \theta^{\rh\ga} \theta^{\mu}_{\pt{\mu}\ga} 
-\et^{\si\nu}\et^{\al\be} \theta^{\rh\ga} \theta^{\mu}_{\pt{\mu}\ga}) 
+ (\mu \rightleftharpoons \nu ) 
\nonumber\\
&& 
\pt{space space space space} 
+ \et^{\rh\nu} \et^{\si\mu} \theta^{\al\ga} \theta^{\be}_{\pt{\be}\ga} 
\}
+\{ \rh \rightleftharpoons \si \} 
- 2 \et^{\rh\si} \et^{\mu\nu} \theta^{\al\ga} \theta^{\be}_{\pt{\be}\ga} 
\big) 
\nonumber\\
&&
+\fr 1{16 \ell^4} C_{(1)} \big( \{ (\et^{\si\nu} \theta^{\rh\al} \theta^{\be\mu} 
+ \et^{\si\nu} \theta^{\rh\be} \theta^{\al\mu} )+(\mu \rightleftharpoons \nu ) \} 
+ \{ \rh \rightleftharpoons \si \} \big)
\quad . 
\label{KNC}
\eea
At this stage one can project \rf{KNC} into the irreducible tensors that appear in \rf{SLVNC}.

Consider the first coefficients, $s^{(4)\mu\rh\al\nu\si\be}$, 
for which the operator it is contracted with, $\sim h \prt \prt h$, 
is a gauge invariant combination (invariant under the transformation $\de h_\mn = -\prt_\mu \xi_\nu - \prt_\nu \xi_\mu$).
Calculation with Young Tableau projection $P_Y$ reveals
\bea
s^{(4)\mu\rh\al\nu\si\be} &=& P_Y^{\scriptsize\young(\mu\nu,\rh\si,\al\be)} (K_{\rm NC})^{\mu\nu\rh\si\al\be} 
\nonumber\\
&=& \frac{1}{36 \ell^4}(2 C_{(1)}-3 C_{(2)}+C_{(4)}+C_{(5)} )
\big( \fr 12 \et^{\rh\si} \theta^{\al\nu} \theta^{\be\mu} - \fr 12 \et^{\nu\rh} \theta^{\al\si} \theta^{\be\mu} 
+ \et^{\rh\si} \theta^{\al\mu} \theta^{\be\nu} + ...\big),
\nonumber\\
\label{s4}
\eea
where the ellipses stand for the remaining symmetrizing terms.
The explicit terms are not shown for brevity and because this contribution 
can be more profitably expressed using an equivalent two-tensor set of coefficients defined by
\beq
{\overline s}_{\ga\de} = -\fr {1}{36} \ep_{\mu\rh\al\ga} \ep_{\nu\si\be\de} s^{(4)\mu\rh\al\nu\si\be}.
\label{sbar1}
\eeq
Employing this, 
the portion of the lagrangian containing the $s^{(4)}$ coefficients can be expressed as
\beq
L_{\rm LV, NC} \supset \fr{1}{4\ka} \int d^4x h_{\mu\nu} {\overline s}_{\ka\la} {\cal G}^{\mu\ka\nu\la},
\label{Lsbar}
\eeq
where, 
for the non-commutative model under study, 
we have
\beq
{\overline s}_{\ka\la} = -\fr {1}{24 \ell^4}(2 C_{(1)}-3 C_{(2)}+C_{(4)}+C_{(5)}) \left( \theta_{\ka\al}\theta_{\la}^{\pt{\la}\al} 
- \fr 14 \et_{\ka\la} \theta^2 \right),
\label{sbar}
\eeq
and we have removed the trace of these coefficients since they contribute only as a scaling of GR at this level.
This result shows that the non-commutative model overlaps with,
in part,
the minimal SME gravity sector in the weak-field limit.
In this model, 
the $9$ coefficients ${\overline s}_{\ka\la}$ are evidently controlled by the $6$ non-commutative parameters $\theta^{\al\be}.$
Note that the size of these coefficients depends on the relative size of the non-commutative parameters 
and the the length parameter $\ell$.

For the other classes of coefficients appearing in \rf{SLVNC} we can proceed in a similar fashion 
with the Young Tableau projection.
All terms are summarized in the table below.  
The explicit expressions for the Young projections are lengthy and omitted here for brevity 
but they can be calculated with standard methods (\cite{hamermesh}).

\begin{table}[H]
\caption{Young Projections for the {\it kinetic} portion of the NC action.}
\label{6classes} 
\centering
\begin{tabular}{cc} 
\vspace{3mm}
SME Coefficients & Young projection \\ 
\vspace{3mm}
$s^{(4)\mu\rh\al\nu\si\be}$ &  $ P_Y^{\scriptsize\young(\mu\nu,\rh\si,\al\be)} (K_{\rm NC})^{\mu\nu\rh\si\al\be}$  \\ 
\vspace{3mm}
$s^{(4,1)\mu\rh\nu\si\al\be}$ &  $P_Y^{\scriptsize\young(\mu\nu\al\be,\rh\si)}(K_{\rm NC})^{\mu\nu\rh\si\al\be} $ \\ 
\vspace{3mm}
$s^{(4,2)\mu\rh\al\nu\si\be} $ &  $P_Y^{\scriptsize\young(\mu\nu\be,\rh\si,\al)}(K_{\rm NC})^{\mu\nu\rh\si\al\be} $ \\ 
\vspace{3mm}
$k^{(4,3)\mu\al\nu\be\rh\si} $ &  $P_Y^{\scriptsize\young(\mu\nu\rh\si,\al\be)}(K_{\rm NC})^{\mu\nu\rh\si\al\be} $ \\ 
\end{tabular} 
\end{table}

\section{Conclusion, Prospects for Further Work} 

We have shown that the model proposed in Ref.~\cite{Ciric:2016isg}, 
in its quadratic limit, 
is a subset of the Lorentz- and diffeomorphism-violating Standard-Model Extension.
The main results are understood as a series of Young Tableau maps described in section $4$.

One consequence of the match obtained relates to experimental and observational constraints
on the noncommutative model considered.
For the gauge-preserving portion of the lagrangian, 
for which the observable effects are controlled by the ${\overline s}_{\mu\nu}$ coefficients, 
an extensive study of phenomenology has been performed \cite{Bailey:2006, Bailey:2009, Bailey:2011, Bailey:2013, Hees:2015}.
To date, 
numerous experiments and observations have reported measurements on 
these coefficients \cite{Kostelecky:2008ts,Hees:2016,Tasson:2016}. 
The best current astrophysical limits come from a recent comparison of the arrival times of electromagnetic and
gravitational waves from a pair of colliding neutron stars \cite{abbot2017}.
Lunar laser ranging and ground-based gravimetry also place limits on these coefficients \cite{llr,gravi,Flowers:2017,Shao:2018}.
\alttextcolor{red}{The best limits imply constraints on the order of ${\overline s}_{\mu\nu}<10^{-14}$.  
Heuristically then, 
this would imply that the non-commutivity coefficients $\theta^{\al\be}$ and the length parameter $\ell$ are related by 
$\theta^2/\ell^4<10^{-15}$.  
However, 
a more precise statement would require a thorough phenomenological analysis 
of the diffeomorphism-violating terms in Section 4 above.}

It would be of interest to explore the role of additional terms in the non-commutative model, 
as in Ref.\ \cite{Dimitrijevic:2014} that involve higher derivatives.
These terms have been generally classified in the SME approach and a match should exist \cite{Kostelecky:2017zob}.

\vspace{6pt} 

\acknowledgments{
Thanks to Berry College 
and the Indiana University Center for Spacetime Symmetries 
for financial support 
during the creation of this work.
Q.~G.~Bailey acknowledges support from the National Science Foundation under Grant No.~PHY-1806871.}

\conflictsofinterest{The author declares no conflict of interest.} 

\appendixtitles{yes} 
\appendixsections{one} 


\appendix

\section{Geometric Quantities to 2nd Order in the Metric Perturbation} 
Consider a pair of theories. 
The first operates in curved-spacetime, 
including a manifold $\cal M$, a metric $g_{\mu\nu}$, 
a local flat metric for tangent spaces $\et_{ab}$, 
and a set of vierbein ${e_\mu}^a$ that relate the metrics 
through $g_{\mu\nu}={e_\mu}^a{e_\nu}^b\et_{ab}$. 
(Equivalently, the vierbein may be thought of as a position-dependent change-of-basis matrix 
that relates a manifold coordinate basis $\{\vec{v}_\mu\}$ 
to a local tangent-space basis $\{\vec{u}_a\}$.) 
The second theory operates in a flat spacetime with an auxiliary field $h_{\mu\nu}$. 
For this theory, 
the manifold is simply $\mathbb{R}^4$, 
the manifold metric is $\et_{\mu\nu}$, 
the tangent-space metric is $\et_{ab}$, 
and global coordinates may be found so that the vierbein is just the Kronecker delta ${\de_\mu}^a$. 

A perturbation scheme is a map 
\beq 
({\cal M},g_{\mu\nu},\et_{ab},{e_\mu}^a) \rightarrow 
(\mathbb{R}^4,\et_{\mu\nu},\et_{ab},{\de_\mu}^a)+h_{\mu\nu} 
\eeq 
between these theories 
so that they approximately describe the same physical effects. 
In particular, 
we will consider situations where $g_{\mu\nu}\approx\et_{\mu\nu}$, 
so that the map may be nicely approximated by a power series 
in $g_{\mu\nu}-\et_{\mu\nu}$. 
We wish to calculate an action in terms of $h_{\mu\nu}$ 
that mimics the physical effects of the original theory 
up to order $h^2$. 

The first piece of the map is defined by the correspondence 
\beq 
g_{\mu\nu} = \et_{\mu\nu} + h_{\mu\nu} 
\quad . 
\label{gethmap} 
\eeq 
This is the definition of $h_{\mu\nu}$ 
and hence is correct to all orders in $h$. 
Our goal in this section is to find expressions for other geometric quantities 
$g^{\mu\nu}$, ${e_\mu}^a$, and so on 
that appear in the action of the full theory. 
The formulas for these quantities should only involve 
the flat-spacetime tensors $h_{\mu\nu}$, $\et_{\mu\nu}$, $\et_{ab}$, and ${\de_\mu}^a$. 

It is important to note that the defining map \Eq{gethmap} 
is not a tensor equation in the original spacetime. 
This implies that indices on $h_{\mu\nu}$ cannot be raised and lowered 
like the indices of true tensors. 
That is, $h^{\mu\nu}$ is {\itshape not} equal to $g^{\mu\al}g^{\nu\be}h_{\al\be}$. 
The geometry of the original manifold does not by itself 
define a unique value of such quantities, 
and we have some freedom in choosing our definition of them. 
The most convenient choice is defining them so that $h_{\mu\nu}$ 
acts like a true tensor in the flat spacetime. 
That is, we pick 
${h^\mu}_\nu := \et^{\mu\al}h_{\al\nu}$, 
$h^{\mu\nu} := \et^{\mu\al}\et^{\nu\be}h_{\al\be}$, etc. 
Similarly, 
we may choose to relate global and tangent-space indices 
with the flat-space veirbein ${\de^\mu}_a$: 
$h_{\mu a}:=h_{\mu\nu}{\de^\nu}_a$, 
${h_\mu}^a:= h_{\mu\nu}\et^{\nu\la}{\de^a}_\la$, etc. 

The raised-index metric $g^{\mu\nu}$ may then be evaluated to second order in $h_{\mu\nu}$ 
through the following strategy. 
The fundamental definition of $g^{\mu\nu}$ is that it is 
the matrix inverse of $g_{\mu\nu}$: 
\beq 
{\de_\mu}^\la = g_{\mu\nu} g^{\nu\la} 
\quad . 
\label{defnupperg} 
\eeq 
We proceed by using the ansatz $g^{\nu\la} = \et^{\nu\la} + j^{\nu\la} + k^{\nu\la} + o(h^3)$ 
where $j^{\nu\la}$ is first order in $h$ and $k^{\nu\la}$ is second order. 
If we insist that \Eq{defnupperg} hold order-by-order in $h$, 
then we need 
\beq 
j^{\al\la} = -\et^{\al\mu}\et^{\nu\la}h_{\mu\nu} 
\quad \mbox{and} \quad 
k^{\al\la} = \et^{\al\mu}\et^{\nu\be}\et^{\la\ga}h_{\mu\nu}h_{\be\ga} 
\quad . 
\eeq 
Using the definitions of upper-index $h$ quantities described in the previous paragraph, 
we may then write 
\beq 
g^{\mu\nu} = \et^{\mu\nu} - h^{\mu\nu} + h^{\mu\al}{h_\al}^\nu + o(h^3) 
\quad . 
\eeq 
Note again that $g^{\mu\nu}\ne \et^{\mu\nu}+h^{\mu\nu}$ 
as the breakdown of $g_{\mu\nu}$ into $\et_{\mu\nu}+h_{\mu\nu}$ 
is not a true tensor operation. 

The quadratic approximation for the vierbein may be calculated 
by using the ansatz ${e_\mu}^a = {\de_\mu}^a + {f_\mu}^a + {\ell_\mu}^a + o(h^3)$, 
where $f$ is first order in $h$ and $\ell$ is second order, 
and insisting that the exact relation 
\beq 
g_{\mu\nu} = {e_\mu}^a {e_\nu}^b \et_{ab} 
\eeq 
hold order-by-order in $h$. 
This results in the expression 
\beq 
{e_\mu}^a = {\de_\mu}^a + \fr{1}{2}{h_\mu}^a - \fr{1}{8}h_{\mu\la}h^{\la a} + o(h^3) 
\quad, 
\eeq 
where again $h$ quantities are related to each other 
with the flat-spacetime metrics $\et_{\mu\nu},\et_{ab}$ 
and flat-spacetime vierbein ${\de_\mu}^a$. 
Explicitly, ${h_\mu}^a:=\et^{\nu\rh}{\de_\rh}^a h_{\mu\nu}$ 
and $h^{\la a}:=\et^{\la\mu}\et^{\nu\rh}{\de_\rh}^a h_{\mu\nu}$. 

Once we have these, 
calculations of other geometric quantities are rather straightforward 
if tedious. 

\begin{description} 
\item[Metric:] 
 \bea 
 g_{\mu\nu} &=& \et_{\mu\nu} + h_{\mu\nu} 
 \quad , 
  \nonumber \\ 
 g^{\mu\nu} &=& \et^{\mu\nu} - h^{\mu\nu} + h^{\mu\al}{h_\al}^\nu + o(h^3) 
 \quad . 
 \eea 
\item[Vierbein:] 
 \bea 
 {e_\mu}^a &=& {\de_\mu}^a + \fr{1}{2}{h_\mu}^a - \fr{1}{8}h_{\mu\la}h^{\la a} + o(h^3) 
 \quad , \nonumber \\ 
 e_{\mu a} &=& \et_{\mu a} + \fr{1}{2}h_{\mu a} - \fr{1}{8}h_{\mu\la}{h^\la}_a + o(h^3) 
 \quad , \nonumber \\ 
 e^{\mu a} &=& \et^{\mu a} - \fr{1}{2}h^{\mu a} + \fr{3}{8}{h^\mu}_\la h^{\la a} + o(h^3) 
 \quad , \nonumber \\ 
 {e^\mu}_a &=& {\de^\mu}_a - \fr{1}{2}{h^\mu}_a + \fr{3}{8}h^{\mu\la}h_{\la a} +o(h^3) 
 \quad , \nonumber \\ 
 e:=\mbox{det}({e_\mu}^a) &=& 1 + \fr{1}{2}{h_\mu}^\mu 
  + \fr{1}{8}( {h_\mu}^\mu{h_\nu}^\nu -2{h_\mu}^\nu{h_\nu}^\mu ) 
  + o(h^3) 
 \quad. 
 \eea 
 Note again that the expressions for the vierbein quantities 
 cannot be related to each other simply by raising and lowering indices: 
 $e^{\mu a}\ne \et^{\mu\la}{e_\la}^a$, etc. 
 Note also that the index placement in the definition of $e$ is important: 
 $\mbox{det}({e^\mu}_a)=\fr{1}{\mbox{det}({e_\mu}^a)}$. 
\item[Connection coefficients:] 
 \bea 
 \Ga_{\al\mu\nu} &=& \fr{1}{2} 
   (\prt_\mu h_{\nu\al} + \prt_\nu h_{\mu\al} - \prt_\al h_{\mu\nu} ) 
 \quad , \nonumber \\ 
 {\Ga^\al}_{\mu\nu} &=& \fr{1}{2} (\et^{\al\si}-h^{\al\si}) 
   (\prt_\mu h_{\nu\si} + \prt_\nu h_{\mu\si} - \prt_\si h_{\mu\nu} ) 
   + o(h^3) 
 \quad , \nonumber \\ 
 {\om_\mu}^{ab} &=& 
   \Big[ -\fr{1}{2}\prt^a {h_\mu}^b 
   +  - \fr{1}{8}h^{a\la}\prt_\mu{h_\la}^b + \fr{1}{4}h^{a\la}\prt_\la{h_\mu}^b 
       - \fr{1}{4}h^{a\la}\prt^b h_{\la\mu} 
   \Big] - \Big[a\rightleftharpoons b\Big] 
   +o(h^3) \nonumber\\
 \quad . 
 \eea 
\item[Derivative compatibility:] 
 \bea 
 \nabla_\ga g_{\mu\nu} &=& 0 
 \quad , \nonumber \\ 
 \nabla_\ga{e_\mu}^a &=& 0 
 \quad . 
 \eea 
\item[Riemann tensor:] 
 \bea 
 R_{\al\be\mu\nu} &=& 
  \Big[ \Big( 
   -\fr{1}{2}\prt_\al \prt_\mu h_{\be\nu} 
   -\fr{1}{8} \prt_\al h_{\mu\la} \prt_\be {h_\nu}^\la 
   -\fr{1}{8} \prt_\mu h_{\al\la} \prt_\nu {h_\be}^\la 
   -\fr{1}{8} \prt_\la h_{\al\mu} \prt^\la h_{\be\nu} 
  \nonumber \\ 
 && \quad 
   -\fr{1}{4} \prt_\al h_{\mu\la} \prt_\nu {h_\be}^\la 
   +\fr{1}{4} \prt_\al h_{\mu\la} \prt^\la h_{\be\nu} 
   +\fr{1}{4} \prt_\mu h_{\al\la} \prt^\la h_{\be\nu}  
  \Big) 
   - \Big(\al\rightleftharpoons\be\Big) \Big] - \Big[\mu\rightleftharpoons\nu\Big] 
   +o(h^3). 
 \nonumber \\ 
 && 
 \eea 
\item[Ricci tensor:] 
 \bea 
 R_{\al\mu}=g^{\be\nu}R_{\al\be\mu\nu} &=& 
   \Big[ 
    \fr{1}{2}\prt_\al \prt_\la {h_\mu}^\la 
    - \fr{1}{4}\prt_\al \prt_\mu {h_\la}^\la 
    - \fr{1}{4}\prt_\la \prt^\la h_{\al\mu} 
   \nonumber \\ 
 && \quad -\fr{1}{2}h^{\la\rh} \Big( 
    \prt_\al \prt_\la h_{\mu\rh} 
    -\fr{1}{2}\prt_\al \prt_\mu h_{\la\rh} 
    -\fr{1}{2}\prt_\la \prt_\rh h_{\al\mu} 
    \Big) 
    \nonumber \\
&& \quad +\Big( 
     \fr{1}{4}\prt_\la {h_\rh}^\rh -\fr{1}{2}\prt^\rh h_{\rh\la} 
     \Big) 
     \Big( 
     \prt_\al {h_\mu}^\la - \fr{1}{2}\prt^\la h_{\al\mu} 
     \Big) 
    \nonumber \\ 
 && \quad 
    -\fr{1}{4}(\prt_\la {h_\mu}^\rh) \Big( 
     \prt_\rh {h_\al}^\la - \fr{1}{2}\prt^\la h_{\al\rh} \Big) 
    +\fr{1}{8}(\prt_\al h_{\la\rh}) 
     (\prt_\mu h^{\la\rh}) 
    \Big] + \Big[\al\rightleftharpoons \mu\Big] +o(h^3). \nonumber\\
 \eea 
\end{description}

\reftitle{References}



\end{document}